\documentclass[review]{elsarticle}
\usepackage{subfigure}
\usepackage{graphicx}
\usepackage{epsfig}
\usepackage{booktabs}
\usepackage[thickspace,amssymb]{SIunits}
\usepackage{lineno}

\begin{document}
\title{Design of a new tracking device for on-line dose monitor in ion therapy}
\author{Giacomo~Traini$^{a,b}$, Giuseppe~Battistoni$^c$, Angela~Bollella$^{a}$, Francesco~Collamati$^{a,b}$, Erika~De Lucia$^d$,
Riccardo~Faccini$^{a,b}$, Fernando~Ferroni$^{a,b}$, Paola Maria~Frallicciardi$^e$, Carlo Mancini-Terracciano$^{a,b}$, 
Michela~Marafini$^{b,f}$, Ilaria~Mattei$^c$, Federico~Miraglia$^{a}$, 
Silvia~Muraro$^c$, Riccardo~Paramatti$^{a,b}$,
Luca~Piersanti$^d$,
Davide~Pinci$^{b}$,
Antoni~Rucinski$^{b,f}$,
Andrea~Russomando$^{a,b}$,
Alessio~Sarti$^{b,f}$,
Adalberto~Sciubba$^{b,f,g}$,
Martina~Senzacqua$^{f}$,
Elena~Solfaroli-Camillocci$^{a,b}$,
Marco~Toppi$^{d}$,
Cecilia~Voena$^{b}$,
Vincenzo~Patera$^{b,f,g}$}
\address{
$^a$ Dipartimento di Fisica, Sapienza Universit\`a di Roma, Pl.e Aldo Moro 2 00185, Roma, Italy; 
$^b$ INFN Sezione di Roma, Pl.e Aldo Moro 2 00185, Roma, Italy; 
$^c$ INFN Sezione di Milano, Via Celoria 16 20133, Milano, Italy;
$^d$ Laboratori Nazionali di Frascati dell'INFN (LNF), Via Enrico Fermi 40 00044, Frascati(Roma), Italy;
$^e$ Istituto di ricerche cliniche Ecomedica, Via Luigi Cherubini 2 50053, Empoli(FI), Italy;
$^f$ Dipartimento di Scienze di Base e Applicate per Ingegneria (SBAI), Sapienza Universit\`a di Roma, Via Antonio Scarpa 14 00161, Roma, Italy;
$^g$ Museo Storico della Fisica e Centro Studi e Ricerche ``E.~Fermi'', P.zza del Viminale 00184, Roma, Italy; 
}
\date{\today}

\begin{abstract}
Charged Particle Therapy is a technique for cancer treatment that exploits hadron beams, mostly protons and carbons. A critical issue is the monitoring of the dose released by the beam to the tumor and to the surrounding tissues. We present the design of a new tracking device for monitoring on-line the dose in ion therapy through the detection of secondary charged particles produced by the beam interactions in the patient tissues. In fact, the charged particle emission shape can be correlated with the spatial dose release and the Bragg peak position. 
The detector uses the information provided by 12 layers of scintillating fibers followed by a plastic scintillator and a small calorimeter made by a pixelated Lutetium Fine Silicate crystal. Simulations have been performed to evaluate the achievable spatial resolution and a possible application of the device for the monitoring of the dose profile in a real treatment is presented.
\\
{\bf Corresponding author}: Cecilia Voena, INFN Roma, P.le Aldo Moro 2, 00185 Roma, Italy, email address: cecilia.voena@roma1.infn.it.
\end{abstract}
\maketitle
%%%%%%%%%%%%%%%%%%%%%%%%%%%%%%%%%%%%%%%%%%%%%%%%%%%%%%%%%%%%%%%%%%%%%%%%%%%%%%
%\linenumbers

\section{Introduction}
Proton and carbon ion beams are presently used to treat selected solid
cancers~\cite{Hadrotherapy}. Compared to the standard X-rays treatments the main advantage of this technique is the better localization of the irradiated dose in the tumor region sparing healthy tissues and possible surrounding organs. This because charged particles loose most of the energy at the end of their path, in the Bragg peak, while X-rays exponentially release their energy with the penetration in matter. Up to now most patients have been treated with proton beams, but use of carbon beams has recently started in Europe.
New dose monitoring devices need to be introduced into clinical use, to fully exploit the capability of particle therapy to deliver the dose as planned  over the cancer position~\cite{faccini,kraan}. Currently beam monitors are used to control the dose application during the patient treatment while some attempts of using PET scans just after the treatment are reported in literature (for a review see~\cite{review_parodi}). 

It has been already demonstrated that the Bragg peak can be correlated with the emission pattern of secondary prompt photons within the 1-10 MeV energy range~\cite{Testa,Catania} and secondary charged particles with kinetic energies up to few hundreds MeV~\cite{Carichi} created by the beam interactions. The carbon beam produces higher energy, more abundant secondary charges with respect to the proton beam. The neutral prompt component is experimentally more challenging but it is fairly produced also in proton treatments.

\begin{figure}[h]
\centering
\includegraphics[width=45mm,height=50mm]{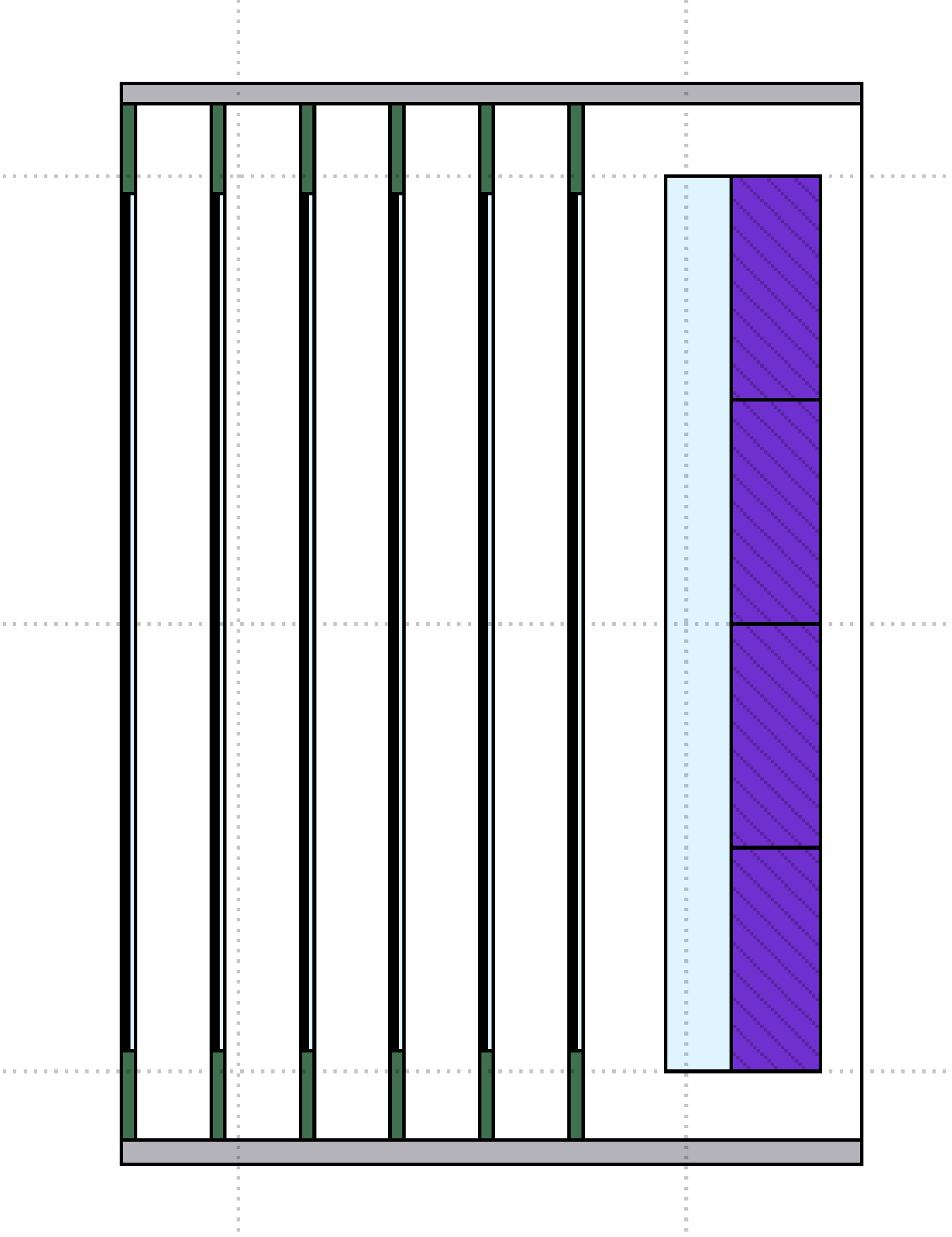}
\includegraphics[width=60mm,height=50mm]{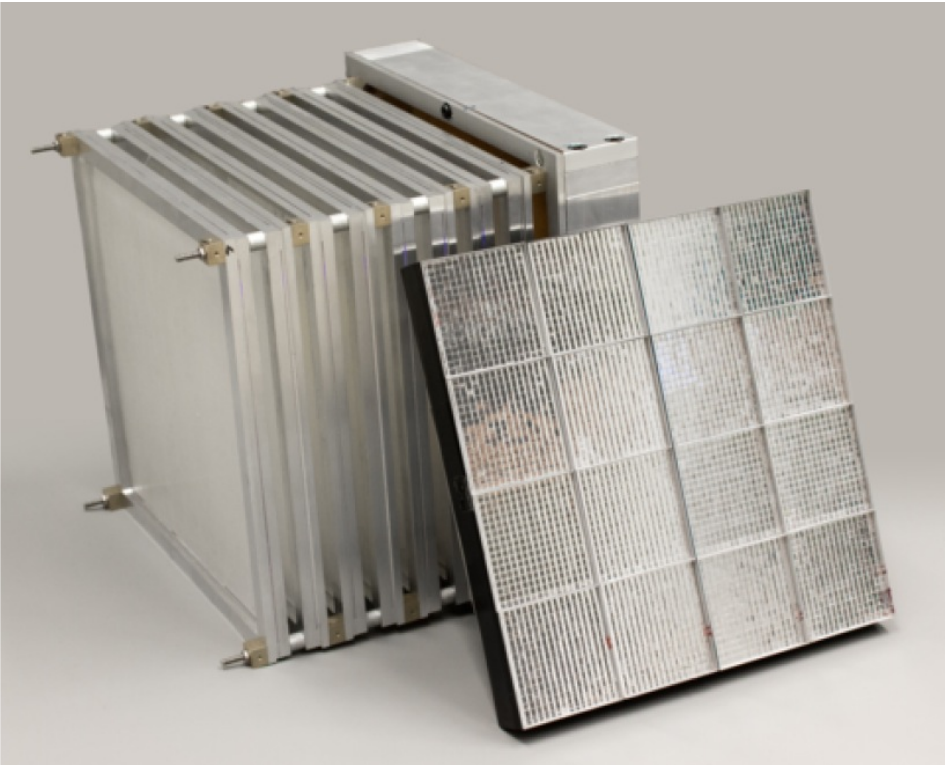}
\caption{Scheme of the detector (left). The six tracking planes (dark lines), the plastic scintillator (light blue) and the LFS calorimeter (violet) are visible. A picture of the detector in the assembling phase (right).}
\label{fig:ProfScheme}
\end{figure}

\begin{figure}[h]
\centering
\includegraphics[width=58mm,height=50mm]{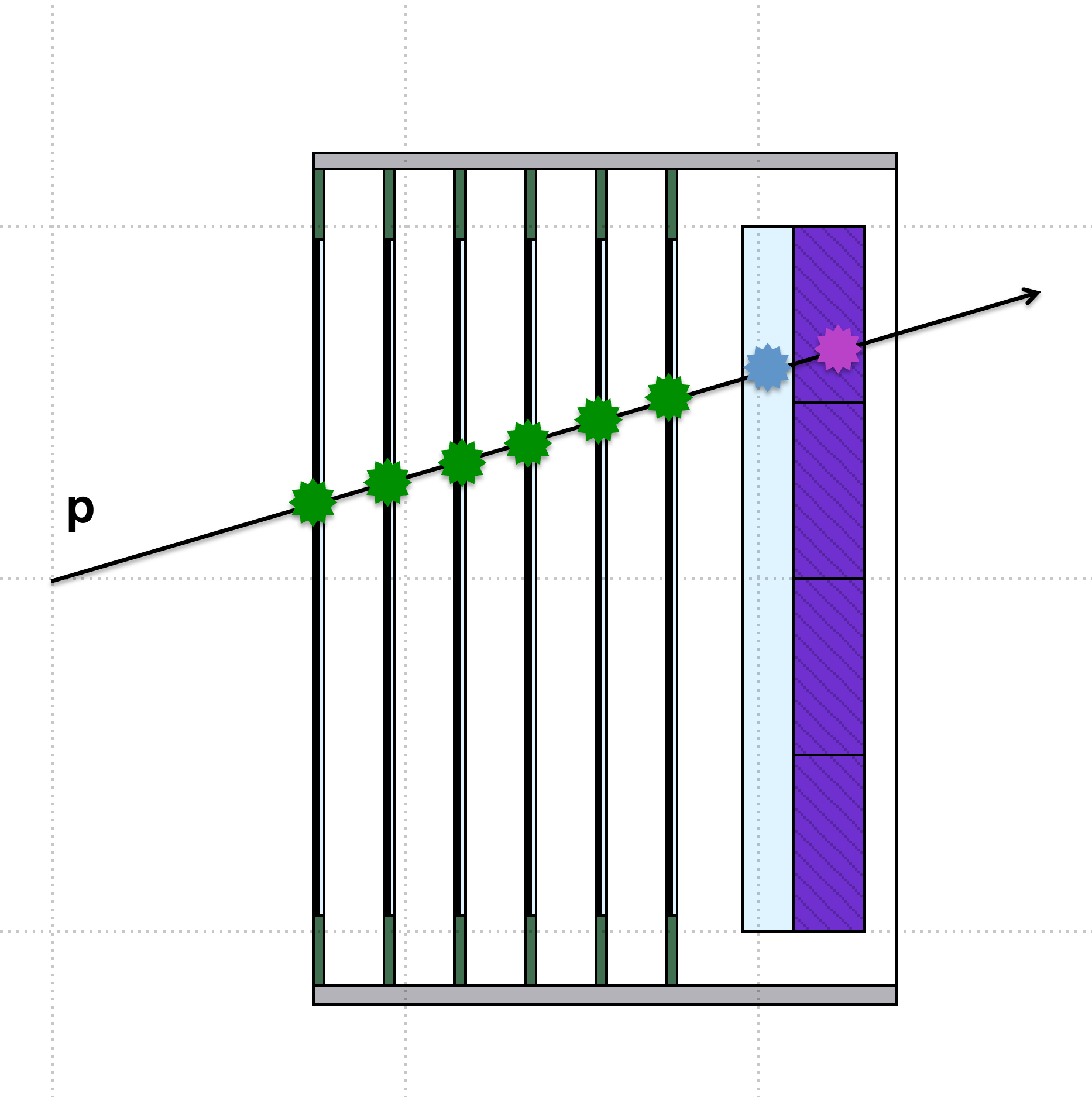}
\includegraphics[width=50mm,height=50mm]{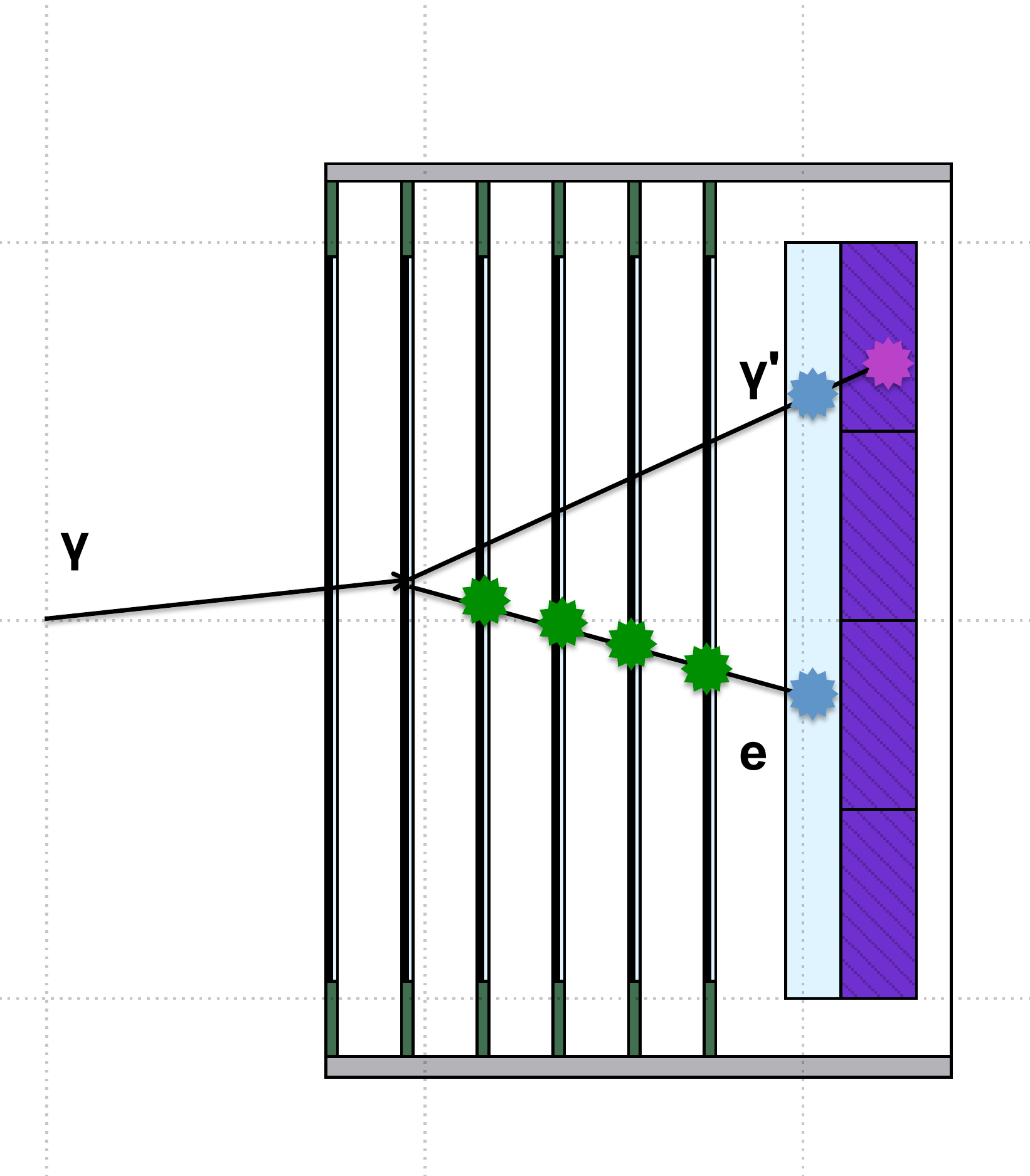}
\caption{Detection principle of a proton (left) and of a photon (through its Compton interaction, right) in the dose profiler.}
\label{fig:Scheme}
\end{figure}

The new device we are proposing has been designed to operate as a charged particle tracker but in addition is capable to reconstruct prompt photons (through Compton interaction or pair production). The detector is composed by 6 planes of orthogonally placed scintillating fiber layers followed by a plastic scintillator and a small calorimeter made of LFS (Lutetium Fine Silicate) crystals. Figure~\ref{fig:ProfScheme} shows a scheme and a picture of the detector. Figure~\ref{fig:Scheme} shows the principle of reconstruction for a charged particle and for a photon (through a Compton interaction) that enter the detector.
The final design is a compromise between compactness, important due to the space limitations in a treatment room, and large geometrical acceptance, which increases the reconstruction efficiency. The amount of material is minimized in order to contain the multiple Coulomb scattering of tracks that limits track angular resolution.
In order to optimize the charged particle detection the choice of the angle of the detector axis with respect to the beam direction is crucial. At narrow angles there is the advantage that the emission flux is enhanced and the charged particle energy is higher (thus the multiple scattering is minimized). On the other hand, due to the projection on the beam line, the spatial resolution on the emission shape worsens and for angles different from 90\degree the emission shape is convoluted with the transverse beam spot size projected on the beam line.
As reported in~\cite{insidePET}, measurements show that the flux of charged particles emitted at large angle from a tissue-equivalent phantom irradiated by a $^{12}$C ions at 220~MeV/u is still relevant: from 1.3~10$^{-2}$ particles/primary at 60\degree \ to 2.7~10$^{-3}$ particles/primary at 90\degree.
We consider two configurations, with the detector axis at 90\degree \ and 60\degree \ with respect to the beam direction. The prompt photon reconstruction is not discussed in this paper.

The dose profiler is part of the INSIDE (INnovative Solutions for In-beam DosimEtry in hadron therapy) project~\cite{insidePennazio,inside}, a multimodal in-beam dose monitor which includes, beside the profiler, also a PET detector and it is designed to operate at CNAO~\cite{cnao}. The PET detector is not described here.

The paper is organized as follows: Section~\ref{secdesign} describes the detector while Section~\ref{Methods} describes the simulation, the event reconstruction algorithms and a method to monitor the dose profile in a real treatment. Section ~\ref{Results} reports the detector performances and the results of the proposed method applied to a specific case. Section~\ref{Discussion} discusses the results and the prospects and Section~\ref{Conclusions} presents the conclusions.

\section{The detector design}~\label{secdesign}
\subsection{The tracker}
The tracker is composed of six planes each made of two orthogonally placed scintillating fiber layers (384 fibers each) to provide bi-dimensional view.

We have adopted scintillating fibers having a square cross section of 500 $\times$ 500 $\mu$m$^2$ (multi-cladding BCF-12 from Saint-Gobain) with the minimal plane separation (2 cm) necessary to accommodate the front-end electronics readout, in order to increase the geometrical acceptance and the compactness of the detector. The choice of the fiber size is the result of an optimization aiming to balance signal amplitude and the total amount of material to be crossed by charged particles.

The fibers readout is performed by means of 1 mm$^2$ Silicon Photomultipliers (SiPM) coupled to the fibers on both the sides. A single layer is read-out by 192 SiPM (96 per side) arranged in a such a way to read-out all the fibers.
In total the sensitive area per layer is 19.2 $\times$  19.2 cm$^2$, read by 192 channels. The SiPM are readout by BASIC32$\_$ADC ASIC~\cite{BASIC}, a custom integrated circuit with  32 analog inputs providing independent voltage offset for the SiPM bias voltage fine adjustment, threshold, gain, shaping, an 8-bit ADC with zero suppression at 20 MHz project frequency, and a fast output for triggering purposes.
On top of the analogical SiPM board a digital board is used to produce the SiPM bias voltage, to distribute the trigger signal and to send data to a further concentrator board. 
The system is designed to sustain a rate of 20 kHz; we expect few kHz of events for a carbon treatment in the detector.

The rate of single photoelectron (p.e.) is 100 kHz per SiPM which, with a BASIC integration time of 100 ns and a threshold of 3 p.e., causes an electronic noise rate per SiPM of 10 Hz. The overall noise is foreseen to be reducible at trigger level to a negligible rate by using the time coincidence of two or more fiber planes and the calorimeter.

The spatial single hit resolution is $\sim$ 300 $\mu$m. A minimum ionizing particle (MIP) interacting in a fiber produces 20 p.e. and an energy resolution of about 25-30$\%$ is expected. A  typical charged particle entering the profiler has an energy from 3 to 20 times a MIP thus we expect a fiber detection efficiency close to $100\%$.

\subsection{The plastic scintillator}
The plastic scintillator (EJ-200 from Eljen) is a polyvinyl-toluene based scintillator with low atomic number ($Z_{eff}$ = 3.4) placed just after the tracker.
It is used, through its energy measurement, for event selection purposes in the reconstruction of charged fragments tracks.
It also prevents electrons from Compton interactions of prompt photons from reaching the calorimeter and from back-scattering in the tracker, thus improving pattern recognition. 

To avoid the development of a dedicated readout subsystem, the front-end board of the tracker is also used to read the scintillator. For this reason this sub-detector is made of 4 independent layers 0.6 cm thick, with the same external dimensions of the fibers frames and with the same 2 cm spacing.
Each layer is made of 16 scintillating slabs with dimensions of 20 cm $\times$ 1.2 cm $\times$ 0.6 cm.
It is expected that this set-up provides about 50 p.e. for each slab per MIP with a 20-25$\%$ energy resolution.

\subsection{The calorimeter}
The role of the inorganic crystal scintillator placed behind the plastic scintillator is to measure the energy release of the particles for trigger and event selection purposes. It can be also used to reconstruct photons.

It is a 64 $\times$ 64 matrix of pixelated LFS crystals arranged in 4 $\times$ 4 blocks of 5 cm $\times$  5 cm $\times$  2 cm (each block is composed by a 16 $\times$ 16 crystal matrix from Hamamatsu). The LFS high atomic number ($Z_{eff}$ = 66) allows a compact design together with a high energy resolution.
The crystal readout is performed by means of Multi Anode Photo-Multiplier (MAPMT H8500 from Hamamatsu: 8 $\times$ 8 anodes, 6.1 $\times$ 6.1 mm$^2$ each).

\section{Methods}\label{Methods}
\subsection{Simulation}\label{secsimu}
The Monte Carlo software used for simulations is FLUKA, release 2011.2~\cite{Fluka1,Fluka2}. Simulated data are organized in ROOT-trees~\cite{Root} mirroring the design of the data acquisition output format. Quenching effects in the scintillators have been implemented in the Monte Carlo according to \cite{Quench}.
The simulation, together with the reconstruction code described in Section~\ref{secreco} have been used to optimize the detector design. The figures of merit considered are the spatial resolution on single protons's origin and the proton reconstruction efficiency. The possibility to use the profiler as a prompt photon detector has also been considered in the optimization.

After the finalization of the detector design, simulations are used to study the achievable performances.
We concentrate on secondary protons produced by carbon ion beams, since protons are the main component of the secondary charged flux.
The beam is simulated with a Gaussian profile in the transverse dimensions with $\sigma$ = 0.8 cm.
The profiler distance from the beam axis is 40 cm.

We use two types of simulations, called ``full simulation'' and ``parametric simulation'' in what follows, used for different purposes.

In the full simulation a $^{12}$C beam of 220 MeV/u impinges on a PMMA (poly-methyl-methacrylicate) cylindrical phantom of 2.5 cm radius. The cylinder length is two times the range of the $^{12}$C primary ions. Two different configurations are studied, with the beam at 90\degree and 60\degree with respect to the detector axis. The fragmentation of the $^{12}$C is simulated by FLUKA.\\
High statistic samples are used to study the event selection, i.e. the discrimination of the particles of interest (secondary protons produced
in the fragmentation of carbons ions) from other particles.

In the parametric simulation the experimental data reported in~\cite{Carichi} are considered, where again a 220MeV/u $^{12}$C  ion beam at 90\degree and 60\degree with respect to the detector axis impinges on a cylindrical PMMA phantom (2.5 cm radius). 
Proton tracks are generated according to the experimental kinematic and spatial distributions.
Several samples have been produced, with different radii of the cylinder (from 2.5 cm to 10 cm) and different primary beam energies (from 112 MeV/u to 220 MeV/u).

\begin{figure}[h!]
\centering
\includegraphics[width=90mm,scale=0.6]{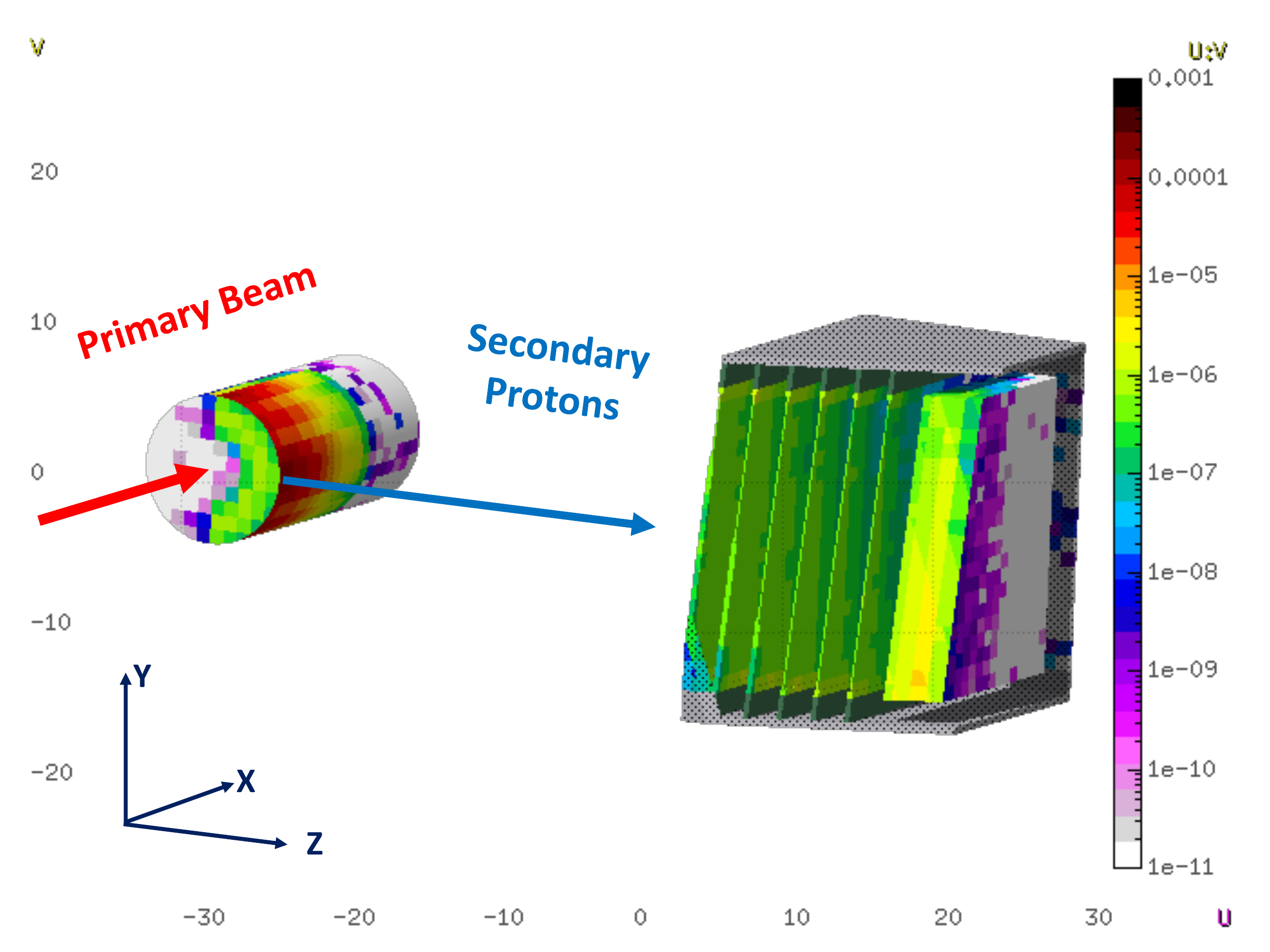}
\caption{Simulated setup described in the text with the carbon ion beam coming along the $x$ direction, at 90\degree with respect to the profiler axis and 2.5 cm radius of the phantom PMMA cylinder.}
\label{fig:simu}
\end{figure}

\begin{figure}[h!]
\centering
\includegraphics[width=80mm,scale=0.6]{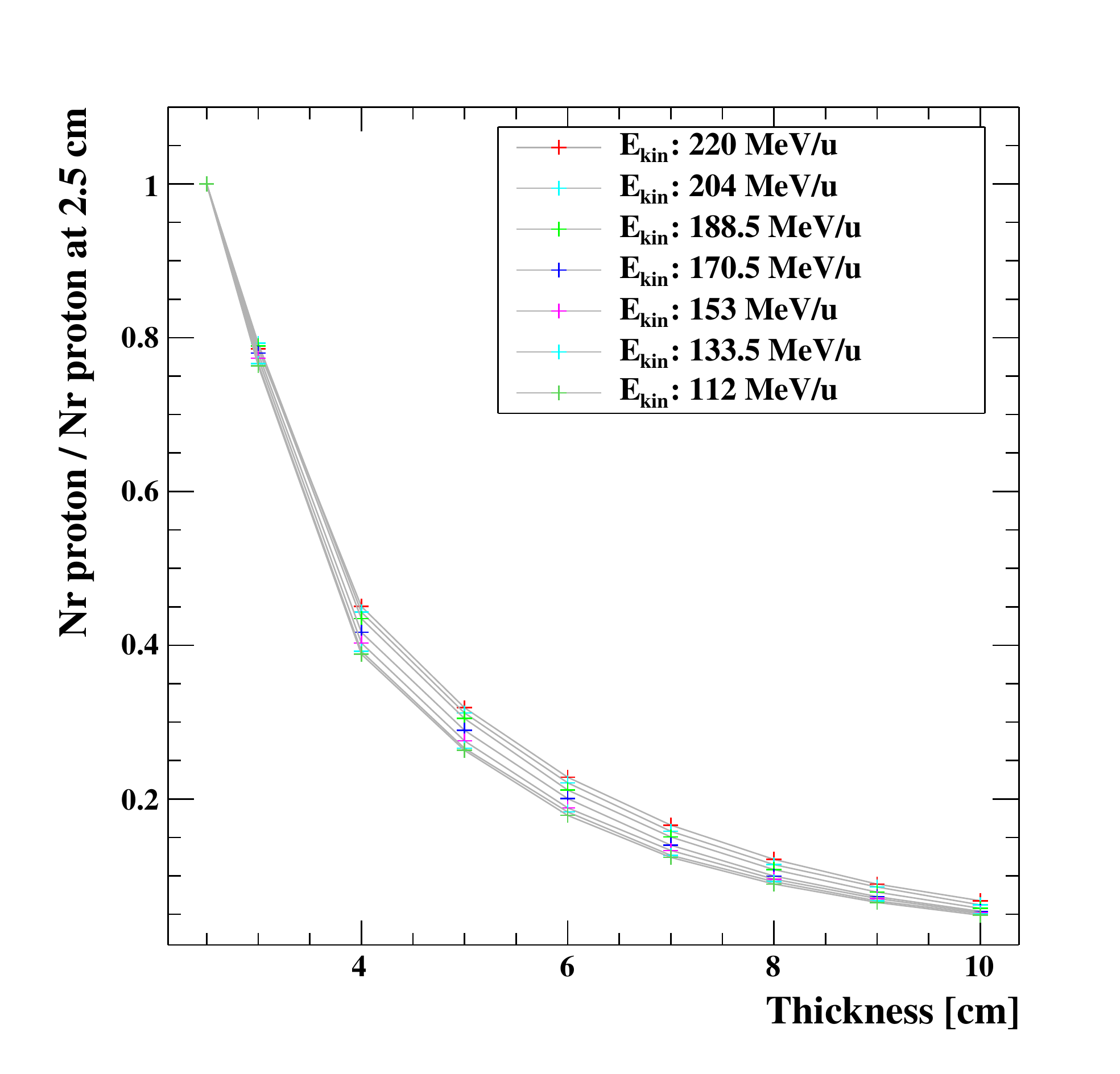}
\caption{Attenuation profile for simulated protons in the PMMA phantom described in the text for different beam energies and different PMMA transversal thicknesses. Different cylindrical phantoms with different radii are used. The carbon ion beam is coming along the $x$ direction, at 90\degree with respect to the profiler axis. The case with the 2.5 cm cylinder radius is taken as a reference.}
\label{fig:attenuation}
\end{figure}

The parametric simulation is used to determine the achievable spatial resolution and to study attenuation effects without relying on FLUKA hadronic models. Moreover, it is used to calibrate the algorithm described in Section~\ref{secreal}.
Figure~\ref{fig:simu} shows a picture of the simulated setup while Figure~\ref{fig:attenuation} shows the attenuation profiles of the protons (i.e. the fraction of protons that survive after a given PMMA thickness) through the different thicknesses of the crossed materials and beam energies, taking as a reference the case with the 2.5 cm cylinder radius. The average kinetic energy of the protons of interest is about 70 MeV. %check
In what follows a reference system where $z$ is the coordinate along the longitudinal detector axis and $x$,$y$ the two transverse coordinates is assumed, as shown in Figure~\ref{fig:simu}.

\subsection{Event reconstruction and analysis}\label{secreco}
The data analysis software has been developed in the programming language C/C$^{++}$ interfaced with ROOT. 
The aim of the reconstruction code is to select events where a secondary proton enters the profiler (Subsection~\ref{evsel}), to reconstruct its track and to determine its origin (Subsection~\ref{reco}).

\subsection{Event selection}\label{evsel}
Protons in the profiler are distinguished on the basis of the energy releases in the different
detectors. The energy release in the tracker, defined as the sum of the energy of all the hits ($E_{fib}$) and that in the scintillator ($E_{scint}$) are studied with Monte Carlo simulation
in order to define energy thresholds for particle identification.
If the energy deposit in the LFS crystal is used in place of that in the scintillator, results similar to those shown below are obtained.

\begin{figure}[h!]
\centering
\includegraphics[width=80mm,scale=0.6]{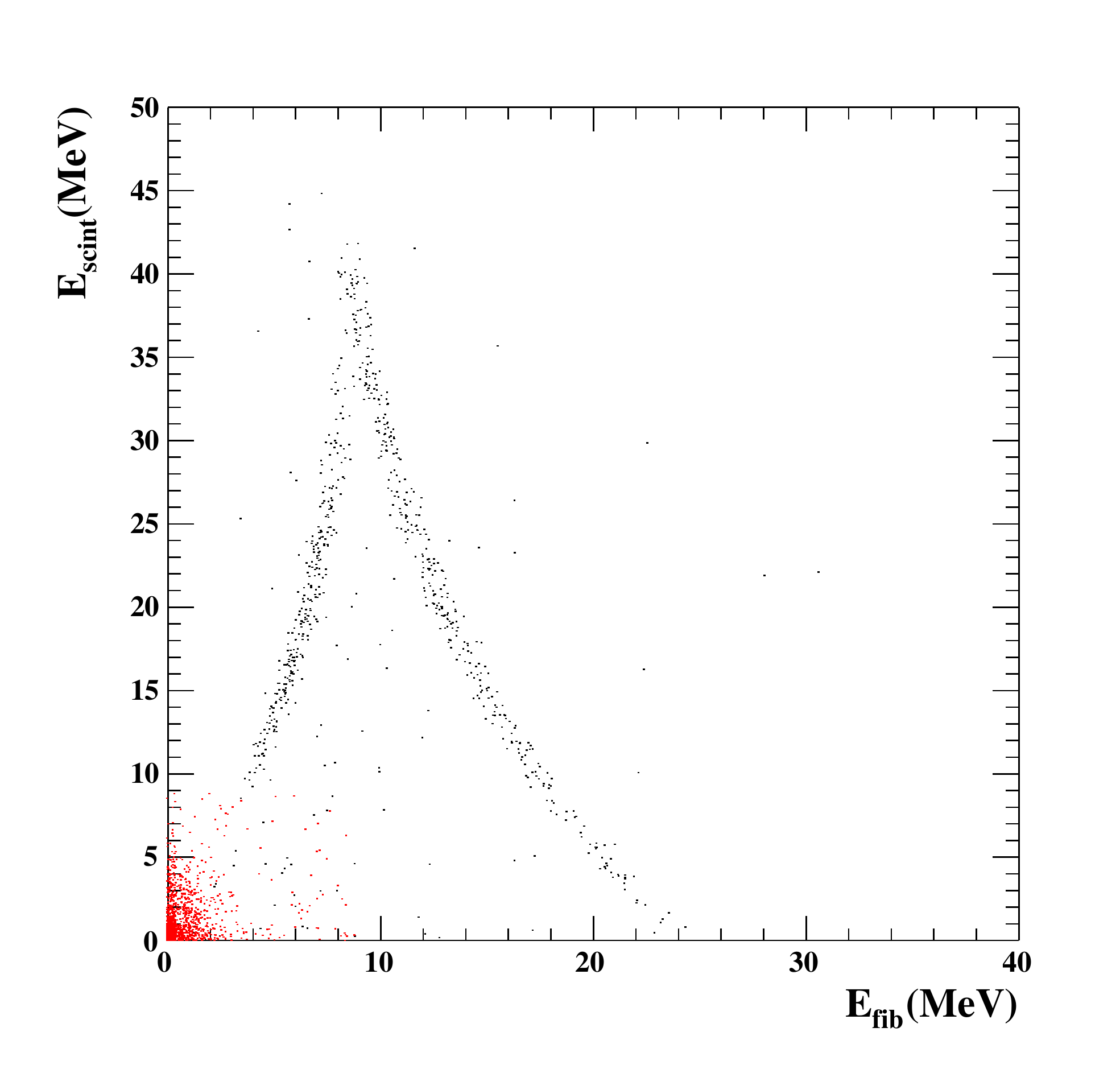}
\caption{Energy deposit for simulated events where a proton track (black dots) or a track from a different particle (red dots) is reconstructed in the profiler.}
\label{fig:evsel1}
\end{figure}

Figure~\ref{fig:evsel1} shows the distribution of $E_{scint}$
as a function of $E_{fib}$ for simulated events with energy releases in both the tracker and in the plastic scintillator where a proton enters the profiler (black dots) and for events with tracks from other particles (red dots). 
Two regions can be distinguished: a region with a two-fold curve, and a region with a low energy deposit in the two detectors ($E_{fib}<$ 4 MeV and $E_{scint}<$ 7 MeV).
The two-fold curve is due to proton tracks reconstructed in the profiler.
The curve can be explained by the fact that protons release energy in the tracker according to the Bethe-Bloch distribution, while in the scintillator two regimes
can be distinguished: a low energy regime where all the energy is deposited in the
scintillator (right curve) and a high energy regime (left curve) where the proton starts to have sufficient energy to go beyond the scintillator and thus also the energy release in this detector starts to follow the Bethe-Bloch curve.
Events outside these curves are tracks due to particle different from protons.

Thus events with $E_{fib}>$ 4 MeV and $E_{scint}>$ 7 MeV are selected as proton events in what follows.
 
\subsection{Proton track reconstruction}\label{reco}
Once that events are selected as proton events, a track is searched in the profiler, starting from the energy deposits (hits) in the fibers.
Fiber hits in the $x$ and $y$ views are clustered separately by grouping hits from consecutive fibers in the same layer to form 2-dimensional clusters ($xz$ and $yz$ views). 
Three-dimensional (3D) clusters are formed by taking all the possible combinations of $x$ and $y$ clusters in a plane. 
A  track finder algorithm is run to build a list of track $seeds$ by grouping consecutive clusters in the first two planes.
Each track $seed$ is prolonged geometrically to subsequent planes and hits are assigned to the different $seeds$ using proximity criteria.
A list of tracks for each event is then available and a $\chi^2$ fit is performed to obtain an estimate of the track parameters (the track is parameterized by two straight lines, in the $xz$ and $yz$ planes). In order to better account for material effects a Kalman filter~\cite{Kalman} is also applied.

In case of events with multiple tracks, the track with the best $\chi^2$ from the fit is chosen. The proton origin in the PMMA is estimated as the geometrical point of closest approach of the track to the beam axis ($x$ axis).

\subsection{Use of the dose profiler in realistic conditions}\label{secreal}
\begin{figure}[h!]
\centering
\includegraphics[width=110mm,scale=0.6]{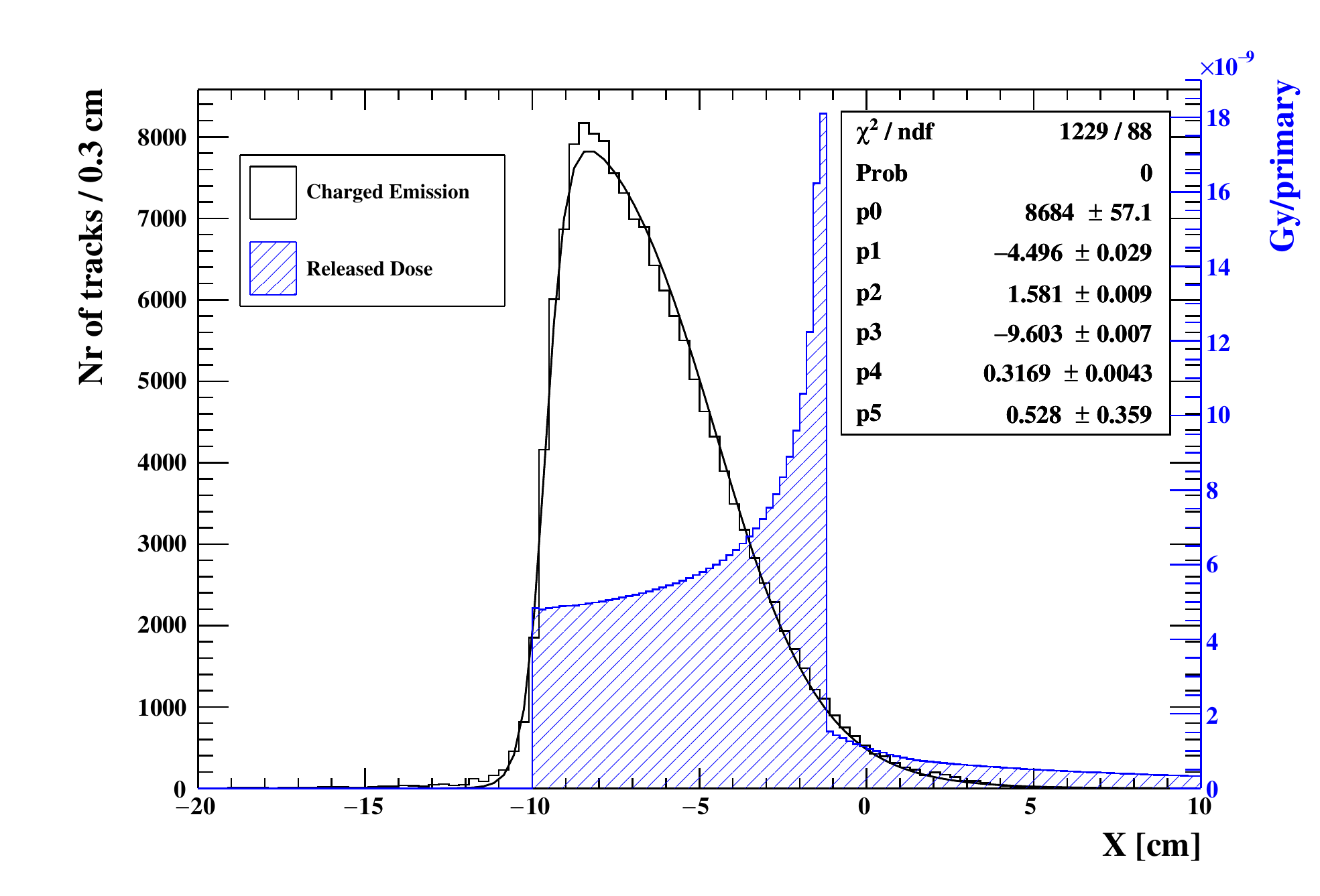}
\caption{Reconstructed secondary charged (protons) emission profile along the beam direction (black histogram) and released dose
(hatched figure) for a carbon ion beam impinging on a PMMA phantom from simulation. Fit (red curve) of the profile with the function of Equation~\ref{DFDfun} is also shown.}
\label{fig:correlation}
\end{figure}

For a carbon ion beam impinging on a cylindrical phantom with 2.5 cm radius, the secondary proton emission profile  along the beam axis can be correlated with the Bragg peak position (see Figure~\ref{fig:correlation}).

This profile, defined as the distribution of the longitudinal coordinate (along the beam direction) of the proton origin $x$, reconstructed as described in Section~\ref{reco}, can be fitted using the function
reported in Equation~\ref{DFDfun}.
\begin{equation}
  f(x) = p_0 \cdot \frac{1}{1+\exp (\frac{x - p_1}{p_2}) } \cdot \frac{1}{1+\exp(-\frac{x - p_3}{p_4}) } + p_5.
  \label{DFDfun}
\end{equation}
Parameters $p_3$ and $p_1$ are related to the rising and falling edge of the distribution, respectively, while $p_4$ and $p_2$ describe the rising and falling slopes of the function. A flat background contribution is modeled through the parameter $p_5$ while $p_0$ is a normalization factor. An example of the above fit can be seen again in Figure~\ref{fig:correlation}. As demonstrated in~\cite{Carichi} in the case of a cylindrical PMMA phantom of 2.5 cm radius, quantities related to the Bragg peak can be computed from the fitted  parameters.

Any complex geometry, like the case of the patient, having different materials, densities and thicknesses, will produce an emission profile which will be quite different from the reference case presented so far. However, since we can get all the relevant information from the patient Computed Tomography (CT), we propose in the following a possible method which allows to take into account all the deformations of the secondary proton emission profile due to the absorption of protons in the patient tissue and to $filter$ $back$ the distribution to the one we are able to correlate directly to the Bragg peak position.

In a real treatment, calibration tables can be produced for different beam energies and $water$ $equivalent$ $material$ thicknesses using simulation. 

\begin{figure}[h!]
\centering
\includegraphics[width=110mm,scale=0.6]{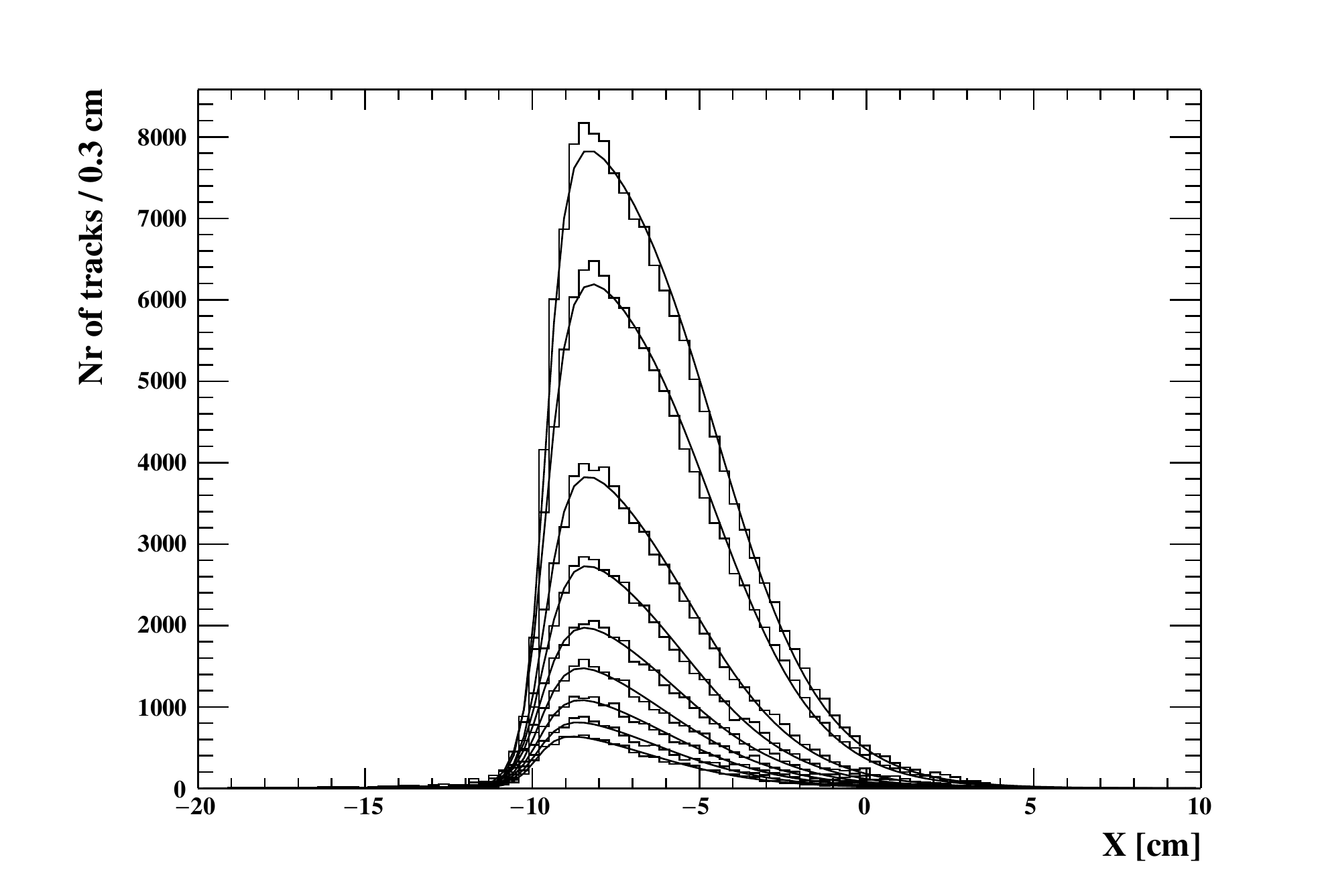}
\caption{Simulation of the reconstructed proton emission profile as detected at $90^o$ with respect to the beam direction, for $^{12}$C beam of 220 MeV/u irradiating a cylindrical PMMA phantom, for different phantom radii. Fits with the function of Equation~\ref{DFDfun} are superimposed.}
\label{fig:Att_BPm10}
\end{figure}

\begin{figure}[h!]
\centering
\includegraphics[width=123mm,scale=0.6]{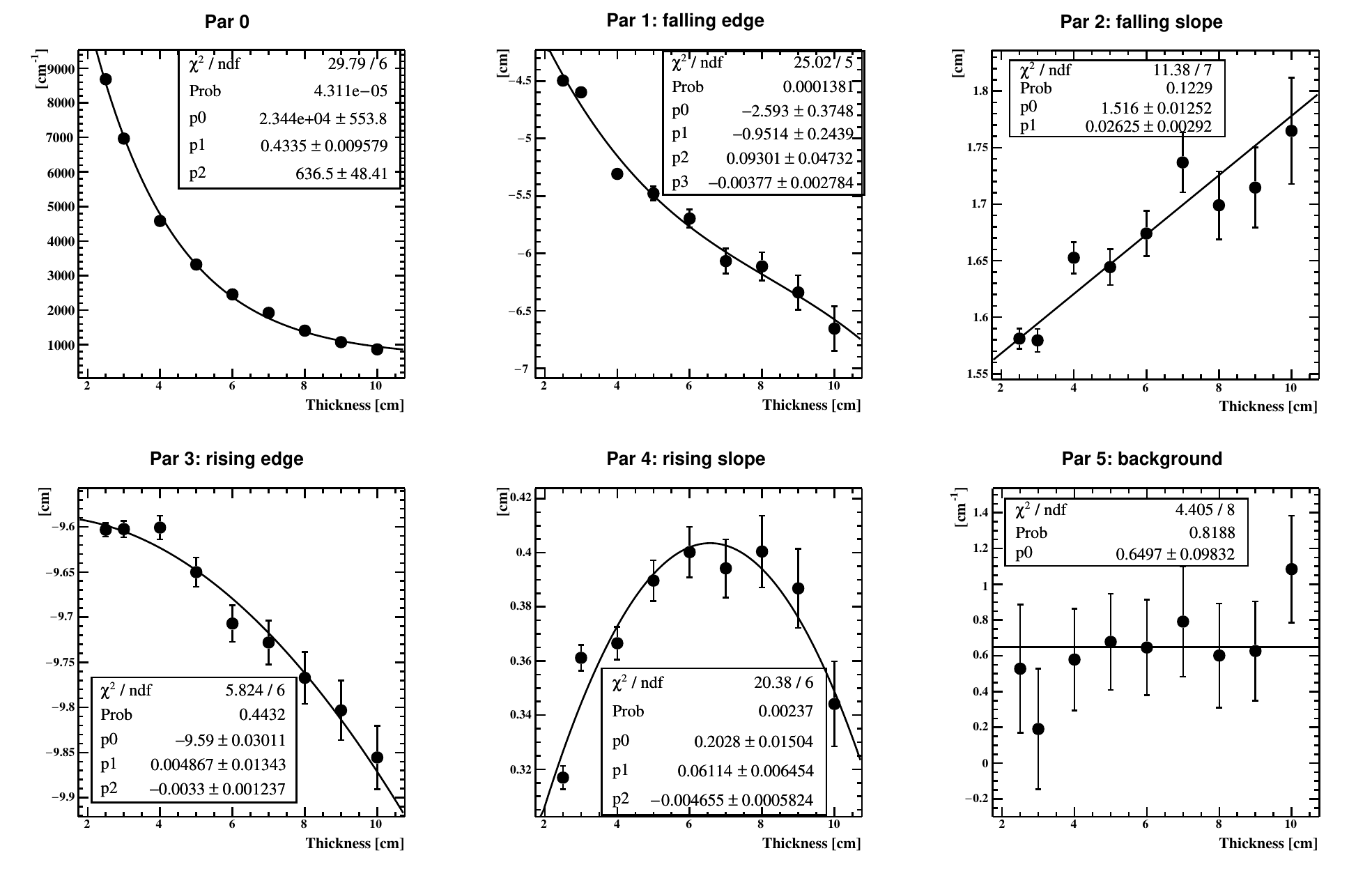}
\caption{Evolution of the fitted parameters $p_i$ (Equation~\ref{DFDfun}) as a function of the PMMA phantom thickness, for 220 MeV/u carbon ion beam (carbon ion beam coming at 90$\degree$ with respect to the profiler axis).}
\label{fig:param}
\end{figure}

The emission profiles predicted by the (parametric) simulation for different thicknesses of the PMMA cylinder are shown in Figure~\ref{fig:Att_BPm10} and are fitted using the function of Equation~\ref{DFDfun}.
The variation of the six $p_i$ parameters is studied as a function of the thickness of the phantom by means of polynomial fits, as shown in Figure~\ref{fig:param}, in the thickness range of 2.5 - 10 cm for $E_{beam}$ = 220 MeV/u. In this way we can parameterize the emission profile for an arbitrary thickness $l$ (material crossed by the secondary protons to exit the phantom).

Parameter $p_{5}$ (background level) is compatible with a constant.
All other parameters exhibit a monotonic behavior except $p_{4}$. This 
parameter is related to the rising slope of the distribution. Due to
multiple scattering contribution increasing with the thickness of 
material, the rising slope is gradually smearing. However the geometrical positioning of the 
detector in our setup limits the possibility of reconstructing tracks 
pointing to a coordinate before the actual
beginning of the phantom ($x$~=~-10~cm in Figures~\ref{fig:correlation}~and~\ref{fig:Att_BPm10}). This artificially
constraints the rising slope to saturate and eventually becoming steeper.

The function of Equation~\ref{DFDfun} can now be generalized as a two variables function of $x$, the emission point along the beam path, and of $l$, the crossed material thickness traversed in the escape path from the phantom, using the $p_{i}(l)$ function:
\begin{equation}
f(x,l) = p_0(l) \cdot \frac{1}{1+exp(\frac{x-p_1(l)}{p_2(l)})} \cdot \frac{1}{1+exp(-\frac{x-p_3(l)}{p_4(l)})} +p_5(l).
\label{eq:2fd_gener}
\end{equation}
For any PMMA cylinder thickness, a weight can be defined for each secondary proton with emission point reconstructed at position $x$ and with crossed material $l$ before escaping the patient:
\begin{equation}
w(x,l) = \frac{f(x,l_0)}{f(x, l)}.
\label{eq:weight}
\end{equation}
Here the reference $l_0$ corresponds to the minimum $2.5$~cm thickness of the PMMA used to collect the data~\cite{Carichi} on which the simulation has been trained. The observed profile for a given thickness can be easily compared after normalization to the reference one.

For materials different from PMMA, to take into account for different densities $\rho_{mat}$ and different material chemical composition, the thickness $l$ is multiplied by the factor
\begin{equation}
F_{mat} = \frac{\rho_{mat} \frac{Z_{mat}}{A_{mat}} }{\rho_{PMMA} \frac{Z_{PMMA}}{A_{PMMA}} }
\label{eq:material_fact}
\end{equation}
where $Z_{mat}$ ($Z_{PMMA}$) is the atomic number and $A_{mat}$ ($A_{PMMA}$) is the mass number of the compound material (of PMMA) respectively.

\section{Results}\label{Results}
\subsection{Detector Performances}\label{secreso}

The relevant quantities that characterize the performances of the detector as a dose profiler are the proton reconstruction efficiency and the resolution on the proton origin along the beam direction ($x$ in our case).

The geometrical acceptance at the considered distance from the beam is about 4$\%$ while the tracking efficiency is close to $100\%$ (given that the fibers are almost fully efficient for protons in the energy range of interest and that noise can be reduced to a negligible level).

The resolution on the proton origin depends on the angular resolution on the track
(i.e the resolution on the track inclination in the $xz$ and $yz$ planes) which in turn depends on the Coulomb multiple scattering (the tracker single-point spatial resolution, 300$\mu$m as stated in Section~\ref{secdesign}, is negligible).

As a reference case, we consider the parametric simulation with the cylindrical PMMA phantom (R = 2.5 cm) with the profiler oriented at 90\degree with respect to the beam direction.
The track angular resolution is about 35 mrad consistent with the MS angle (the dominant MS is that inside the phantom).

\begin{figure}[h!]
\centering
\includegraphics[width=80mm,scale=0.6]{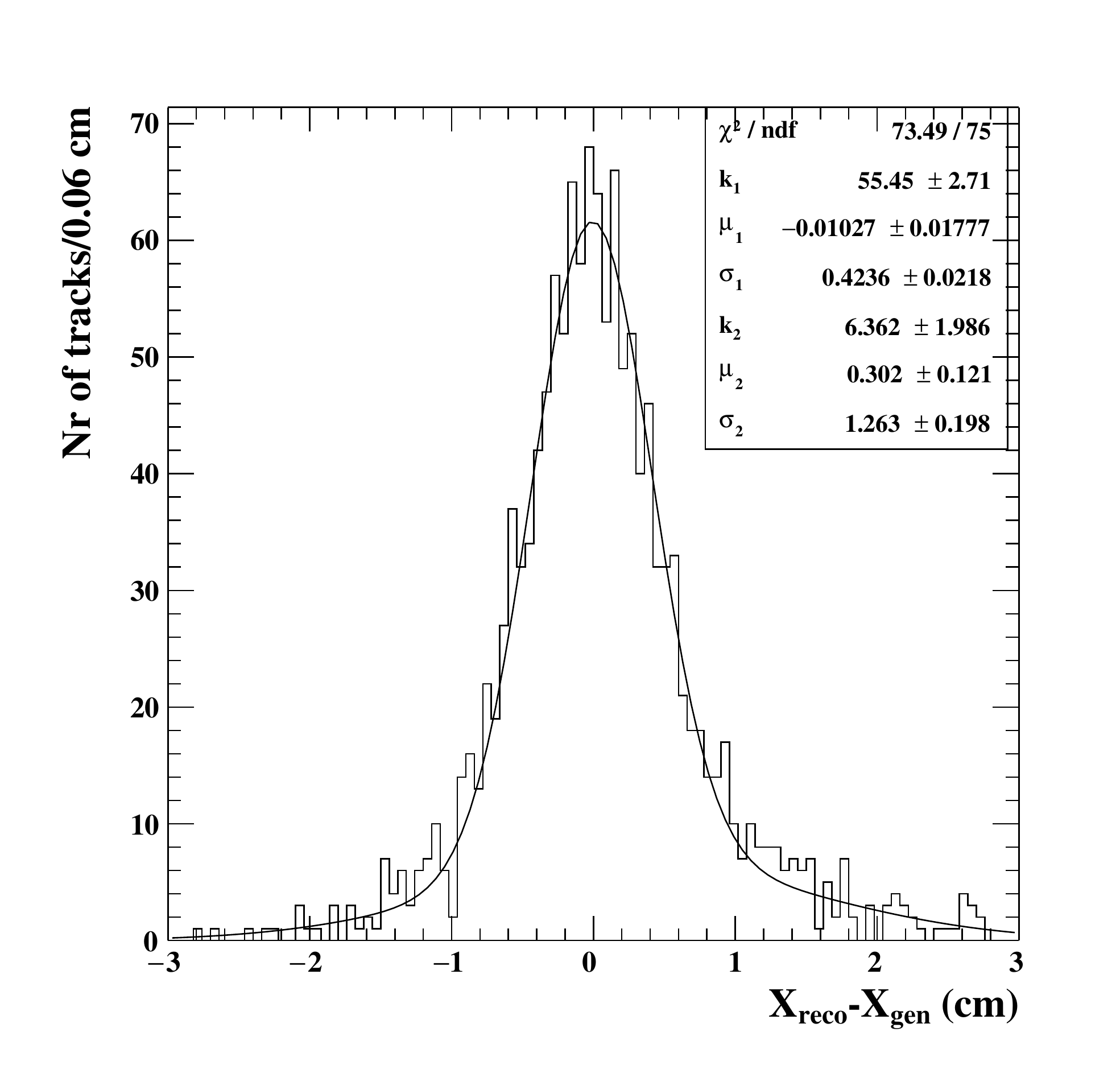}
\caption{Distribution of the difference between the measured and the true value of the proton origin coordinate $x$ (beam direction), in the simulation (carbon ion beam coming along the x direction, at 90\degree with respect to the profiler axis, 2.5 cm radius of the phantom PMMA cylinder). A double Gaussian fit is superimposed to give an estimate of the corresponding resolution.}
\label{fig:XYZreso}
\end{figure}

Figure~\ref{fig:XYZreso} shows the difference between the reconstructed and the true (i.e. generated) $x$ coordinate of the proton origin, fitted to the sum of two Gaussian functions. The width of the distribution is $\sigma_X$ = 0.42 $\pm$ 0.02 cm for 75$\%$ of the events; this can be considered as an estimate of the spatial resolution of the profiler along the beam direction.
The resolutions in the other coordinates, i.e. in the transverse plane with respect to the beam, are $\sigma_Y \simeq 0.4$~cm and $\sigma_Z \simeq 0.3$~cm; a geometrical cut on these coordinate can be possibly applied on data to remove background or mis-reconstructed events.
The configuration with the profiler at 60\degree with respect to the beam axis (again with the PMMA cylinder radius of 2.5 cm) has also been considered.
In this case the resolution on the proton origin along the beam is worse due to the inclination of the beam axis with respect to the profiler ($\sigma_X$ $\sim$ 0.6 cm in this case).

It should be remarked that these resolutions refer to a single track. The goal information on the beam range is extracted from the entire emission shape distribution. The final accuracy on the Bragg peak estimated position will be also function of the number of secondaries detected during the treatment.

\subsection{Use of the dose profiler: an example}

\begin{figure}[h!]
\centering
\includegraphics[width=80mm,scale=1.]{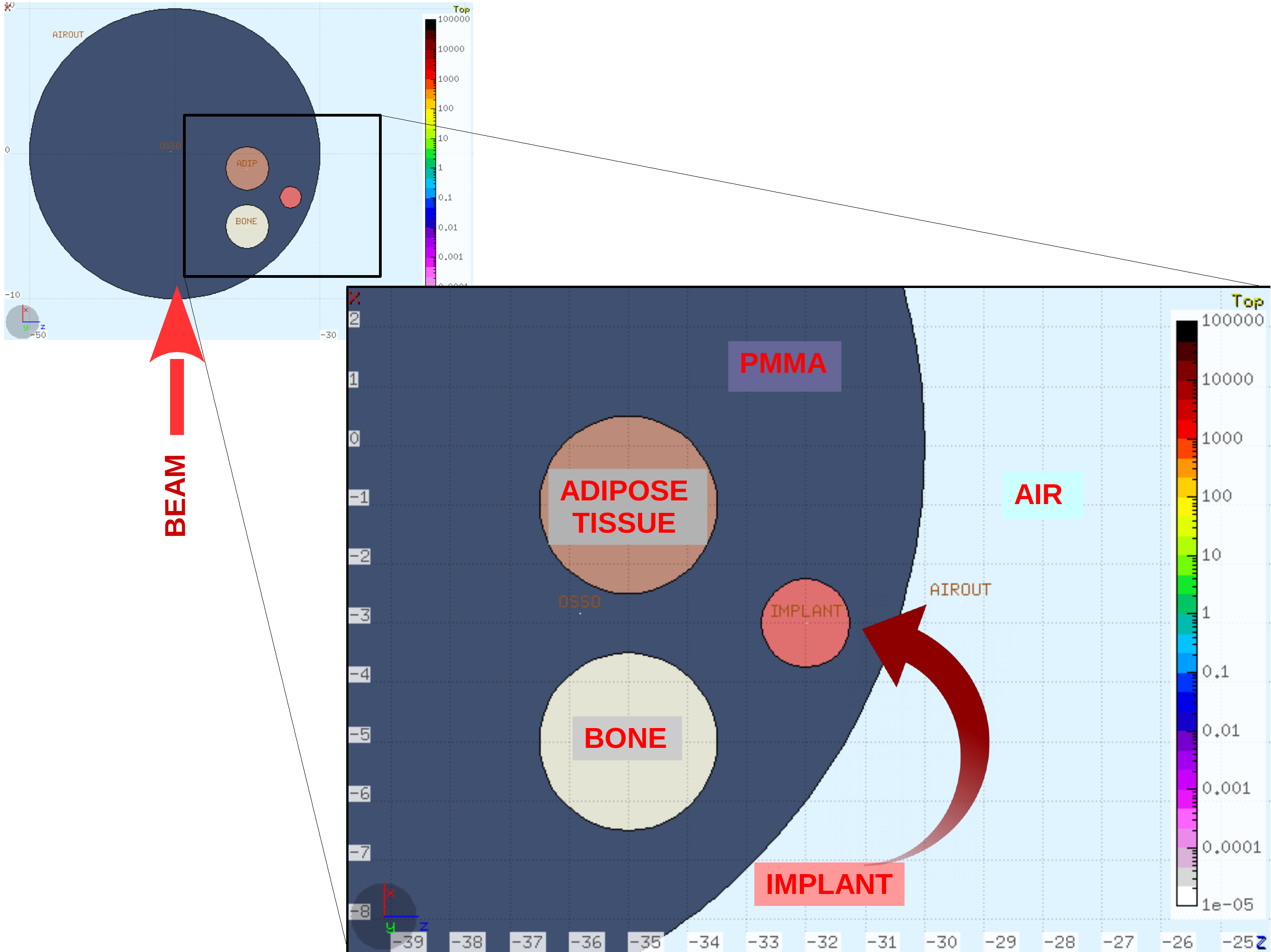}
\caption{PMMA sphere containing three smaller spheres of materials composed as described in the text.}
\label{fig:results1}
\end{figure}

As an example, we show the results of the procedure described in Section~\ref{secreal} using as patient phantom a PMMA sphere containing three smaller spheres of different materials shown in Figure~\ref{fig:results1}. The material crossed by the secondary charged particles from the beam axis to the dose profiler in this sphere is compatible with the one crossed during a head treatment. The number of primary $^{12}$C ions considered in the simulation is the one necessary to cover 1 cm$^2$ area , in the transverse plane with respect to the beam direction, of the distal slice (at 220 MeV/u) of a raster scanning treatment plan delivering 1 Gy of physical dose in 3$\times$3$\times$3~cm$^3$ in water starting at a depth of 7 cm from the skin.\\
In this test we used the composition of three materials which are assigned, in the parameterization of~\cite{Schneider,Parodi2007a,Parodi2007b}, to adipose tissue, bone and metallic implant (Table~\ref{tab:material}).\\
\begin{table}
\centering  
  \small
\begin{tabular}{lccc}
  \toprule
Material    &   Density       & Z/A   & Average Excitation  \\
            &   [$g/cm^2$]    &       & energy [eV]  \\ 
\midrule
PMMA             &   1.190    & 0.539 & \ 74.00  \\ 
Adipose tissue   &   0.926    & 0.557 & \ 63.22  \\  
Bone             &   1.816    & 0.517 & 104.05  \\  
Metallic implant &   2.466    & 0.482 & 107.67  \\ 
\end{tabular}
\\[0.5cm]
\begin{tabular}{lcccccccccc}
\toprule
Material         &   \multicolumn{10}{c} {Chemical composition ($\%$)}         \\
                 &    H   &  C   &  N   &  O   &  Na  &  P   &  S   &  Cl  &  Mg  &  Ca  \\ 
\midrule
PMMA             &   53.3 & 33.3 &      & 13.3 &      &      &      &      &      &      \\  
Adipose tissue   &   11.6 & 68.1 &  0.2 & 19.8 &  0.1 &      &  0.1 &  0.1 &      &      \\  
Bone             &   \ 3.9 & 17.9 &  4.1 & 42.9 &  0.1 & \ \ 9.6 &  0.3 &      &  0.2 & 21.0 \\  
Metallic implant &   \ 3.4 & 15.5 &  4.2 & 43.5 &  0.1 & 10.3 &  0.3 &      &  0.2 & 22.5 \\ 
\end{tabular}
\caption{Material properties and composition as parameterized in~\cite{Schneider,Parodi2007a,Parodi2007b}}
\label{tab:material}
\end{table}
These parameterizations are often used in simulations to assign an elemental composition and a density starting from the patient CT scan.

The calibration curves are obtained as described in Section~\ref{secreal} for the different materials and the expected emission profile has been normalized to the reference one using the weights of Equation~\ref{eq:weight} and the factor of Equation~\ref{eq:material_fact}.
\begin{figure}[h!]
\centering
\includegraphics[width=105mm,scale=0.6]{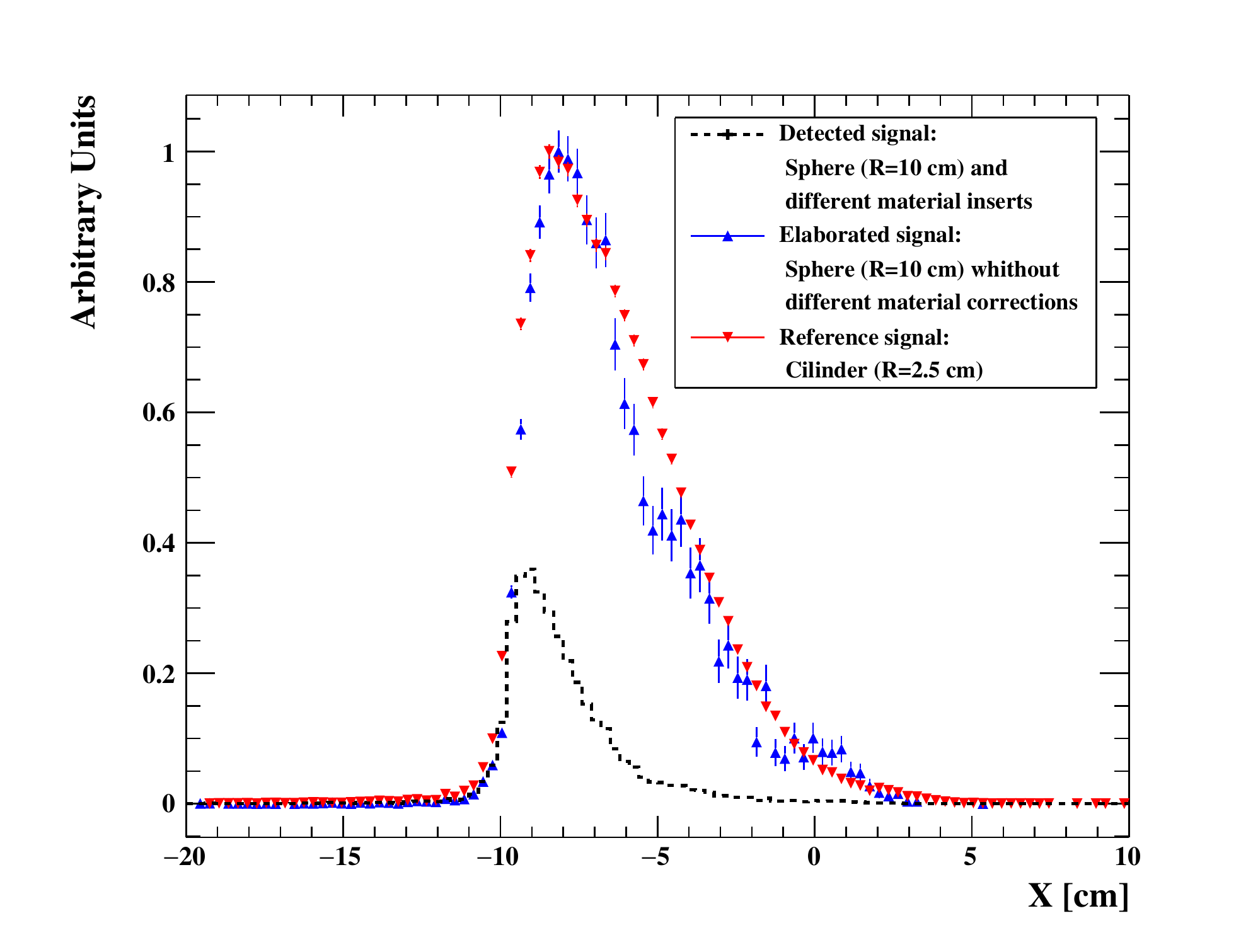}
\includegraphics[width=105mm,scale=0.6]{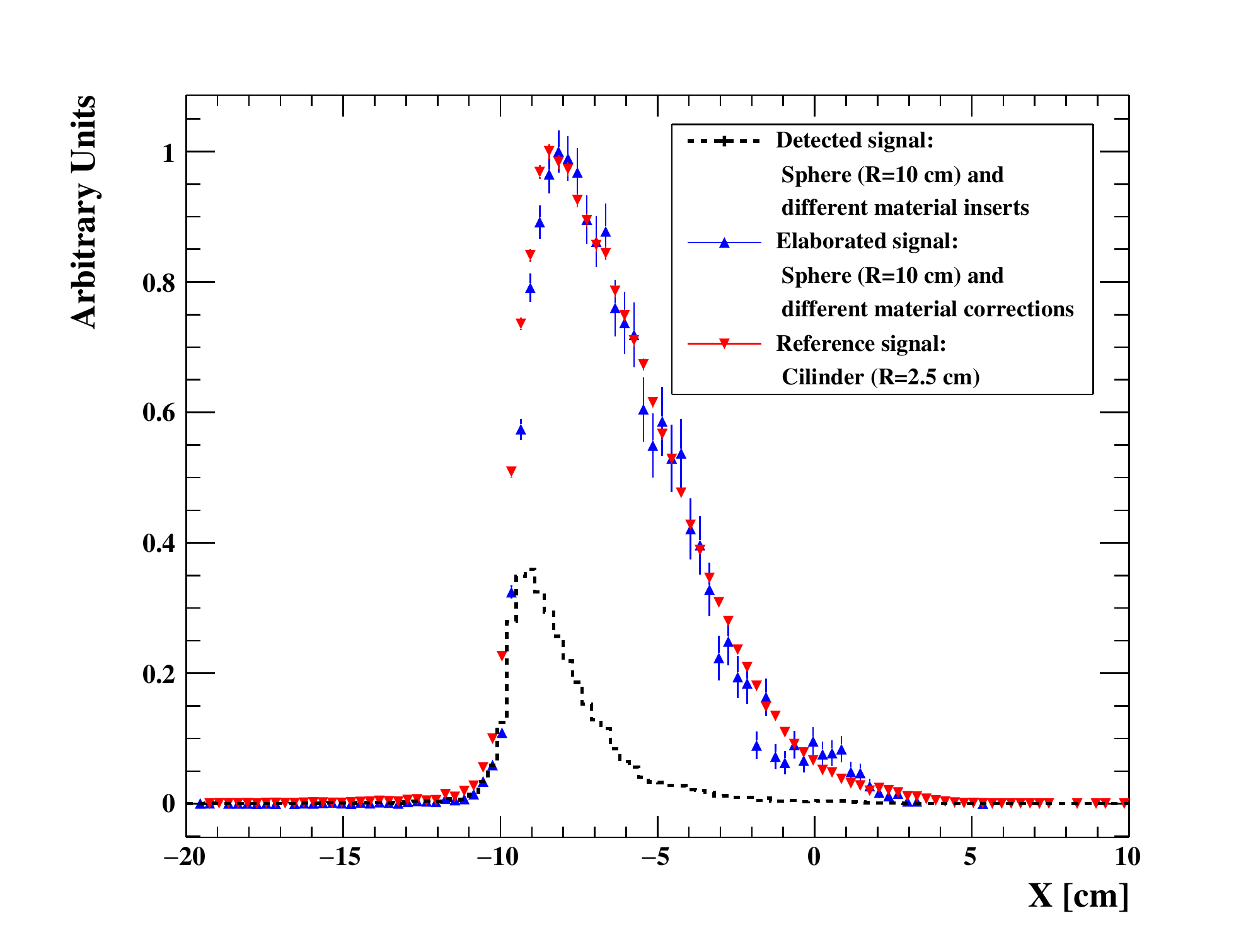}
\caption{Reconstructed reference signal (PMMA cylinder with 2.5 cm radius) ({\it{red}}),  reconstructed signal ({\it{black}}) and back-filtered signal ({\it{blue}}) from a PMMA sphere containing three smaller spheres of different materials for the simulation described in the text. Top: the signal is elaborated by applying only the thickness correction of Equation~\ref{eq:weight}. Bottom: the signal is elaborated taking into account also the different material correction factor of Equation~\ref{eq:material_fact}.}
\label{fig:results2}
\end{figure}
By applying only the thickness correction of Equation~\ref{eq:weight}, a mismatch between the back-filtered and the reference profile still remains (Figure~\ref{fig:results2} top).
Instead, it can be seen that taking into account also the material correction factor of Equation~\ref{eq:material_fact}, the back-filtered sphere distribution and the reference distribution nicely agree (Figure~\ref{fig:results2} bottom). The last step 
to monitor the beam range position is to exploit the correlation between the back-filtered emission profile and the Bragg peak position. A correlation method has been presented in~\cite{Carichi} providing a typical accuracy of 0.3 cm achievable in case of a 220 MeV/u carbon beam in PMMA.

\section{Discussion}\label{Discussion}

The example above shows the feasibility of the prediction of the charged secondaries (mostly protons) emission profile from the patient in a carbon ion treatment, once the detailed map of the material crossed by the detected protons is known (from CT).

The proposed technique can be used in ion therapy with active scanning and allows the monitoring of the Bragg peak position. It could also be used to provide information related to the patient positioning. 

In a real treatment we propose the following procedure that allows
to compare the expected and the released dose on-line:
\begin{itemize}
\item During the treatment each particle detected by the dose profiler is associated to the direction and position of the primary pencil beam delivered at that time. This information is made available on-line by the beam delivery system of the ion therapy facility.
\item By means of a fast reconstruction code, the emission profile is measured. The actual geometry and material are obtained from the patient CT available before the treatment.
\item The proper weight is calculated for each proton according to the procedure described in Section~\ref{secreal}.
\item The resulting profile is compared to the reference one (the 2.5 radius PMMA cylinder in our case) in order to point out possible mismatches between the Bragg peak actual position and the expected one.
\end{itemize}

This procedure will be object of an experimental validation campaign to be performed in the next future.

\section{Conclusions}\label{Conclusions}
We designed a new-concept device to measure on-line the dose pattern in a ion therapy treatment, exploiting the charged tracks, mainly protons, coming out
from the patient during the treatment. We performed a full simulation and event reconstruction  in the detector in order to study the achievable resolution.
We developed a way to measure the dose profile and compare it with expectation.
This detector, particularly suited for carbon ion beams, is under construction and will be tested at CNAO in view of its integration in the multimodal monitoring system designed by the INSIDE collaboration.

\begin{center}\textbf{\large Acknowledgements}\end{center}{\large \par}
This work was supported by $Ministero$ $dell'$ $Istruzione$, $dell'$ $Universit\grave a$ $e$ $della$ $Ricerca$ of the Italian government (PRIN 2010-2011 project nr. 2010P98A75).\\
We also thank Marco Magi (SBAI) and the electronic service at LNF for the profiler design and realization.
%%%%%%%%%%%%%%%%%%%%%%%%%%%%%%%%%%%%%%%%%%%%%%%%%%%%%%%%%%%%%%%%%%%%%%%%%%%%%%

%

%%%%%%%%%%%%%%%%%%%%%%%%%%%%%%%%%%%%%%%%%%%%%%%%%%%%%%%%%%%%%%%%%%%%%%%%%%%%%%
\end{document}